\documentstyle[aps,prl,multicol,fancyheadings,graphicx]{revtex}
\begin{document}

%\draft

\title{Hierarchical organization of cities and nations}

\author{G. Malescio$^1$, N. V. Dokholyan$^{2,3}$, S. V. Buldyrev$^2$,
and H. Eugene Stanley$^2$}

\address{
$^1$Dipartimento di Fisica, Universita' di Messina\\
and\\
Istituto Nazionale Fisica della Materia\\
98166 Messina, Italy\\
$^2$Center for Polymer Studies and Department of Physics\\
Boston University, Boston, MA 02215 USA\\
$^3$Department of Chemistry and Chemical Biology\\
Harvard University, 12 Oxford Street, Cambridge, MA 02138 USA}

\date{10 May 2000}

\maketitle

%fax: 617 353 3783
\begin{multicols}{2}

\noindent {\bf Universality in the behavior of complex systems often
reveals itself in the form of scale-invariant distributions that are
essentially independent of the details of the microscopic dynamics. A
representative paradigm of complex behavior in nature is cooperative
evolution. The interaction of individuals gives rise to a wide variety
of collective phenomena that strongly differ from individual
dynamics---such as demographic evolution, cultural and technological
development, and economic activity. A striking example of such
cooperative phenomena is the formation of urban aggregates \cite{bl1,bl2}
which, in turn, can be considered to cooperate in giving rise to
nations. We find that population and area distributions of nations
follow an inverse power-law behavior, as is known for
cities \cite{z,mhs1,mhs2}.  The exponents, however, are radically
different in the two cases ($\mu \approx 1$ for nations, $\mu \approx 2$
for cities). We interpret these findings by developing growth models for
cities and for nations related to basic properties of partition of the
plane. These models allow one to understand the empirical findings
without resort to the introduction of complex socio-economic factors.}

The way in which urban aggregates are distributed was first investigated
by Zipf  \cite{z} who, half a century ago, observed that the population
distribution of cities follows a power-law behavior with exponent $\mu
\approx 2$. This Zipf law has a ``universal'' character since it holds
at the world level as well as within a single nation, and the exponent
is essentially independent of the area of the nation and its
socio-economical conditions. More recently it has been observed
 \cite{mhs1,mhs2} that the area distribution of satellite cities, towns
and villages around huge urban centers also obeys a power-law with
exponent $\mu\approx 2$.

The remarkable universal result $\mu\approx 2$ has very recently
attracted the attention of a number of physicists who model urban
growth processes \cite{mhs1,mhs2,zm,mz,mmz}. Makse et
al.~ \cite{mhs1,mhs2}, using a correlated percolation model in the
presence of a density gradient, reproduced the observed morphology of
cities and the area distribution of sub-clusters in an urban
system. Zanette and Manrubia  \cite{zm} proposed a stochastic model which
generates intermittent spatiotemporal structures, and predicts a
population distribution in agreement with that observed
empirically. Very recently Marsili and Zhang  \cite{mz} proposed a model
based on a master equation approach which is able to give a population
distribution close to that obeyed by cities. The transition
probabilities which enter the master equation were related to some
estimate of the interactions among individuals living in the same city.
Though the concept of interaction among human beings is not so clearly
defined as that among particles, it is nevertheless true that
individuals living in the same city are related to each other by a
number of ``links'' which in the end define the very concept of city.

People living within the same nation are also related to each
other---e.g., they share language and cultural heritage. Since the
average ``interactions'' among the inhabitants of a nation may differ
from those among people living in the same city, one could ask if the
distribution of nations obeys the same law as that of cities.  In order
to answer to this question we analysed the population and area
distributions of the world's nations. The log-log plot of the population
distribution $f(P)$ is shown in Fig.~\ref{fig:1}a for all nations of the
world. In the same figure we show also the population distribution for
the 140 largest city agglomerates of the USA. Both distributions obey a
power law dependence, $f(P) \sim P^{-\mu_P}$. Regression fits give
$\mu_P=0.97$ for nations, and $\mu_P=1.94$ for
cities. Figure~\ref{fig:1}b shows a log-log plot of the area
distribution $f(A)$ for all nations of the world as well as for nations
belonging to a subset of all nations (Europe).  In both cases $f(A)$
obeys a power law dependence, $f(A) \sim A^{-\mu_A}$ with exponent close
to unity.

The strikingly different values of the exponent for cities and nations
suggest fundamental differences in the historical and social processes
that lead to such distributions.  Here we propose a simple explanation
of the distributions of nations and cities based only on geometrical
considerations. We observe that, at least in principle, there is no
restriction to the land accessible to a nation except that of the total
existing land: a sufficiently powerful nation could expand to absorb all
other nations. Cities, being the result of spontaneous aggregation of
individuals around sites having attractive features, can form away from
existing ones. This separates the plane region into land basins, and
each new city spans a single basin.  The resulting distribution of areas
is not destroyed when a city expands to absorb nearby cities and gives
rise to a compact urban aggregate.  In fact, unlike nations, cities
usually do not lose their land to neighbours: small towns and villages
retain their identity and usually become administrative districts of the
bigger aggregate (obeying, as shown in Ref.~\cite{mhs1,mhs2}, the same
area distribution that holds for separate cities). Thus a major
difference in the way according to which land is occupied by cities and
nations is that for cities the accessible land is fragmented into basins
while it is not for nations.

Next we model the land occupation processes of nations and cities. We
first note that the above considerations suggest that geometric
properties may be behind the differences between city and nation
distributions.  Hence we analyse these processes as random partitionings
of the plane.  Since different nations (or cities) do not all form at
the same time, we consider partition processes in which the different
portions are sequentially selected \cite{rbo,Mekjian97}.

One of the simplest ways of partitioning a plane is to divide it using
straight lines that are randomly oriented and positioned. Each line
divides the region into two portions, of which the smaller is selected
and the larger is further partitioned. This process is close in spirit
to the way in which land is occupied by nations. Because this partition
model resembles the positioning of fences, we refer to it as fence
model.
 
This new model can be solved analytically.  After $n$ partitions, the
land ``available'' for further division is $A_n$, and $A_n=r_nA_{n-1}$,
where $r_n$ is a random factor uniformly distributed between 1/2 and 1.
The area $T_n$ of the ``taken'' portion is $T_{n}=(1-r_{n})A_{n-1}$, and
its logarithm can be written as $\ln T_n=\sum_{i=1}^{n-1}\ln
r_i+\ln(1-r_n)+\ln A_0$.  If we plot on the $x$ axis the values of $\ln
T_n$, we get random points, the average distance between which is equal
to $-\langle\ln r_i\rangle=1-\ln 2\approx 0.3$. The largest nation
corresponds to almost an entire continent $A_0/2\ell$, and the smallest
one corresponds to $A_0(1/2\ell) e^{-0.3n}$, where $n$ is the number of
nations on the continent, and the factor $2\ell$ corresponds to the
average $\langle\ln(1-r_4)\approx -1.7$.  Thus, the distribution of the
logarithm of a nation's area in each continent is a flat distribution
between $\ln A_0-0.3n^{-1.7}$ and $\ln A_0^{-1.7}$ described by the
probability density $P(\ln S)=1/0.3n$.

The world has 5 continents; some, such as Australia and North America,
have very few nations and some, such as Europe and Africa, have
many. Because there are, on average, approximately $n=50$ nations per
continent, we can expect the approximate distribution of the logarithms
of the number of nations to be uniformly distributed between the average
largest country $\sim 10^3$km$^2$ and $10^7\cdot e^{-15}\approx
3$km$^2$, which is consistent with a spread observed in the distribution
of real nations.  The flat distribution of the logarithms $const$
$d(\log S)$ corresponds to the distribution $(const/S)$ $dS$ of the
areas, which is close to what we observe in Fig.~1b. Note, however,
that, due to the few nations in each bin, a particular realization of
the partition process described above may significantly deviate from the
flat distribution we would expect.

The above model could be, however, oversimplified. This concern can be
alleviated by incorporating into it the possibility that nations can evolve,
growing or shrinking. Consider, e.~g., the variant of the fence model 
(called, in the following, evolution model) in which with probability $1/2$ 
a nation can grow or shrink by some given amount. The $P(A)$ histogram, 
remarkably, is not affected.  To see this, we first note that such change 
in area corresponds to the variable $\log(A)$ increasing or decreasing by 
a constant number, i.e., a simple random walk in the variable $\log(A)$.  
As we mention above, $P(A) \sim 1/A$ is equivalent to $P[\log(A)] \sim const$.

In the present case, the random walk is confined by reflecting
boundaries: $A_{\mbox{\scriptsize max}}=A_0$ is some minimal size
$A_{\mbox{\scriptsize min}}$ below which the nation is not stable.  The
distribution of a random walk confined in an interval with reflecting
boundaries converges to a uniform distribution~ \cite{weiss}.  Thus
$P[\log(A)]$ converges to a constant and hence $P(A)\sim 1/A$. The the
$P(A)$ distribution is immune to the ``noise'' of growth and shrinking.
The distribution of nation areas, simulated using the evolution model,
is shown in Fig.~1c. The distribution of the logarithms of the nation
areas are shown in Fig.~1d.

To model city distributions, we shall consider a radically different way
of partitioning the plane. As suggested by the way urban geographers
have thought about ``central place'' theory and the hierarchy of
towns \cite{chlo1,chlo2}, we assume that cities have on the average a
circular shape and thus can be approximately represented as circles.
Land occupation by cities can thus be modeled through the partition of
the plane in nonoverlapping circles. Each new portion is a circle (with
radius chosen from a uniform distribution) at a randomly-chosen
position, but outside of previously-selected circles. The maximal area
of each new circle is limited by the distance from the closest existing
circle. The resulting fragmentation of accessible land reduces the space
available for the next circle more rapidly with respect to what happens
for the portions generated through the fence model, thus making small
portions much more probable. Hence we expect $\mu_A$ to be larger for
the present model.

We simulate the circle model and find that the distribution of the
circle areas follows a power-law behavior which is close to the
empirical data of Fig.1a. A linear fit of the distribution (with the 
exclusion of the region affected by finite size effects) gives an exponent 
$\mu\approx 1.94$ (Fig.1e).  One advantage of the circle model, compared to 
the previous models \cite{mhs1,mhs2,zm,mz}, is that it is based only on the 
geometrical features of the land occupation process.

Another advantage of this model is that it can be solved analytically in
the limit when the area of the newly-formed circle is a small number
proportional to the area of the unoccupied land closest to the center of
this circle, with proportionality coefficient $k \ll 1$. Suppose that
there are $N$ circles in a region of total area $A_0$. Then the
probability that a randomly chosen point is surrounded by an empty space
of area larger than $X$ is governed by a Poisson distribution $F(X,N) =
\exp{(-X N/A_0)}$.  The number of circles with area between $A$ and
$A+dA$ is $p(A,N)dA = N/(A_0 k)\exp{(-AN/A_0 k)}dA$, the derivative of
the Poisson distribution function. The distribution of circles, $P(A)$,
at some given instant when there are $N_c$ circles is given by the
integral $\int_0^{N_c}p(A,N)dN = (kA_0/A^2)[1-\exp(-A N_c/A_0 k)]$.  For
large $N_c$, we recover $P(A)\sim A^{-2}$, so $\mu_A=2$, in agreement
with empirical findings. It is interesting to note that the exponent
$\mu_A =2$ is robust since, as our simulations show, it holds even in
the case when $k$ is not a small number, but is any random number
constrained only by the fact that two circles cannot overlap (as for
urban agglomerates that consist of tightly-packed villages and towns).

In summary, we show, through the analysis of accurate geographical and
demographical data, that the nation population and nation area
distributions obey power-laws. The exponent is, however, surprising and
completely unexpected. Moreover, its value (1) is quite different from
the known exponent for cities (2).  In addition to our empirical
discovery, we propose an explanation for why the population distribution
exponent of cities differs so strikingly from that of nations.

We thank J. S. Andrade Jr., M. Batty, R. S. Dokholyan, I. Grosse, P. L.
Krapivsky, H. A. Makse, M. Morrissey, and S. Redner for interesting and
stimulating discussions. We thank NSF for support. N. V. D. is supported
by NIH postdoctoral fellowship (GM20251-01).

\newpage

\begin{figure}
%\begin{minipage}{\dimen0}
\begin{center}
\includegraphics[width=6.0cm]{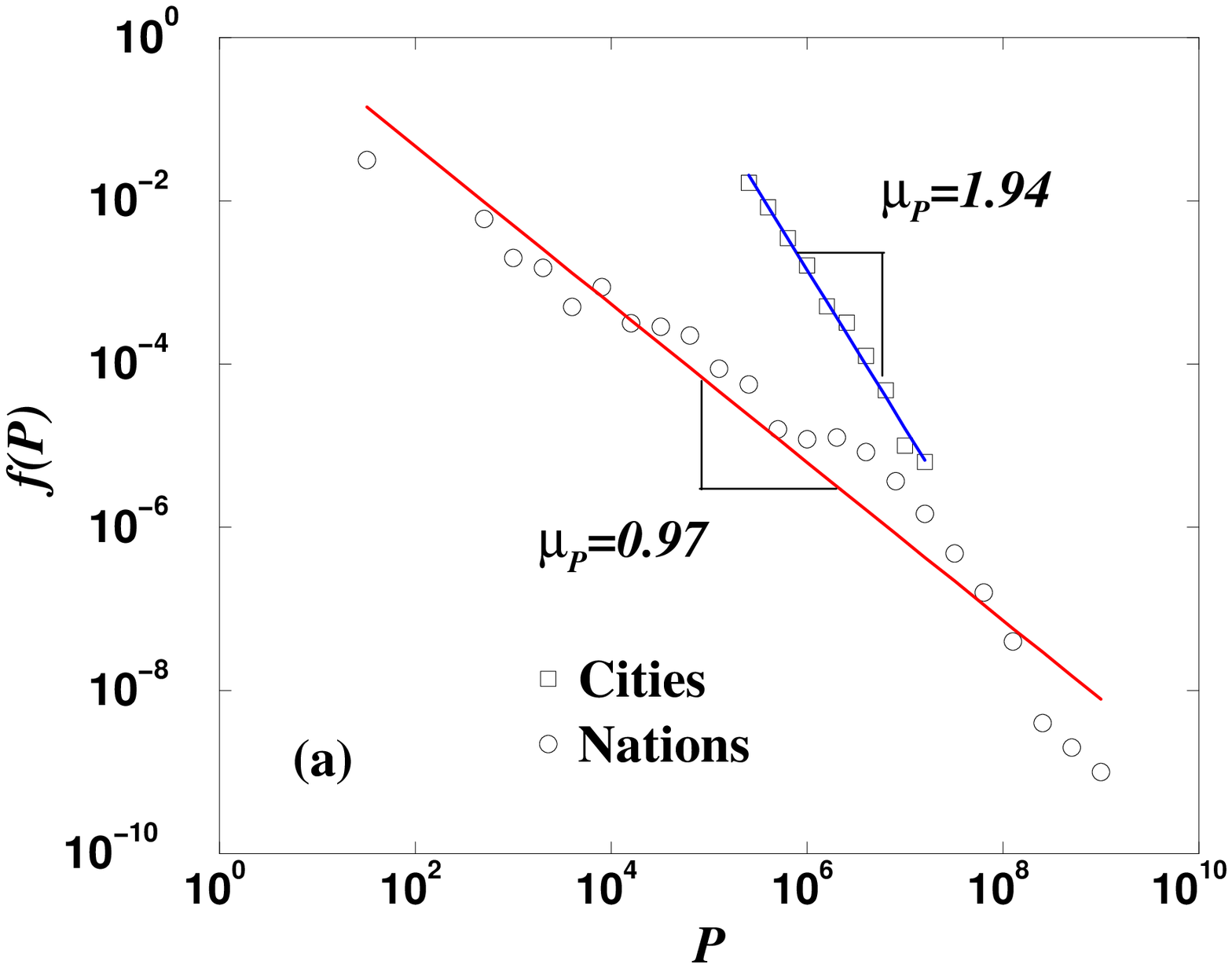}
\end{center}
%\end{minipage}
%\mbox{ \epsfxsize=10cm \epsffile{ fig1a.eps } }
\end{figure}

\begin{figure}
%\begin{minipage}{\dimen0}
\begin{center}
\includegraphics[width=6.0cm]{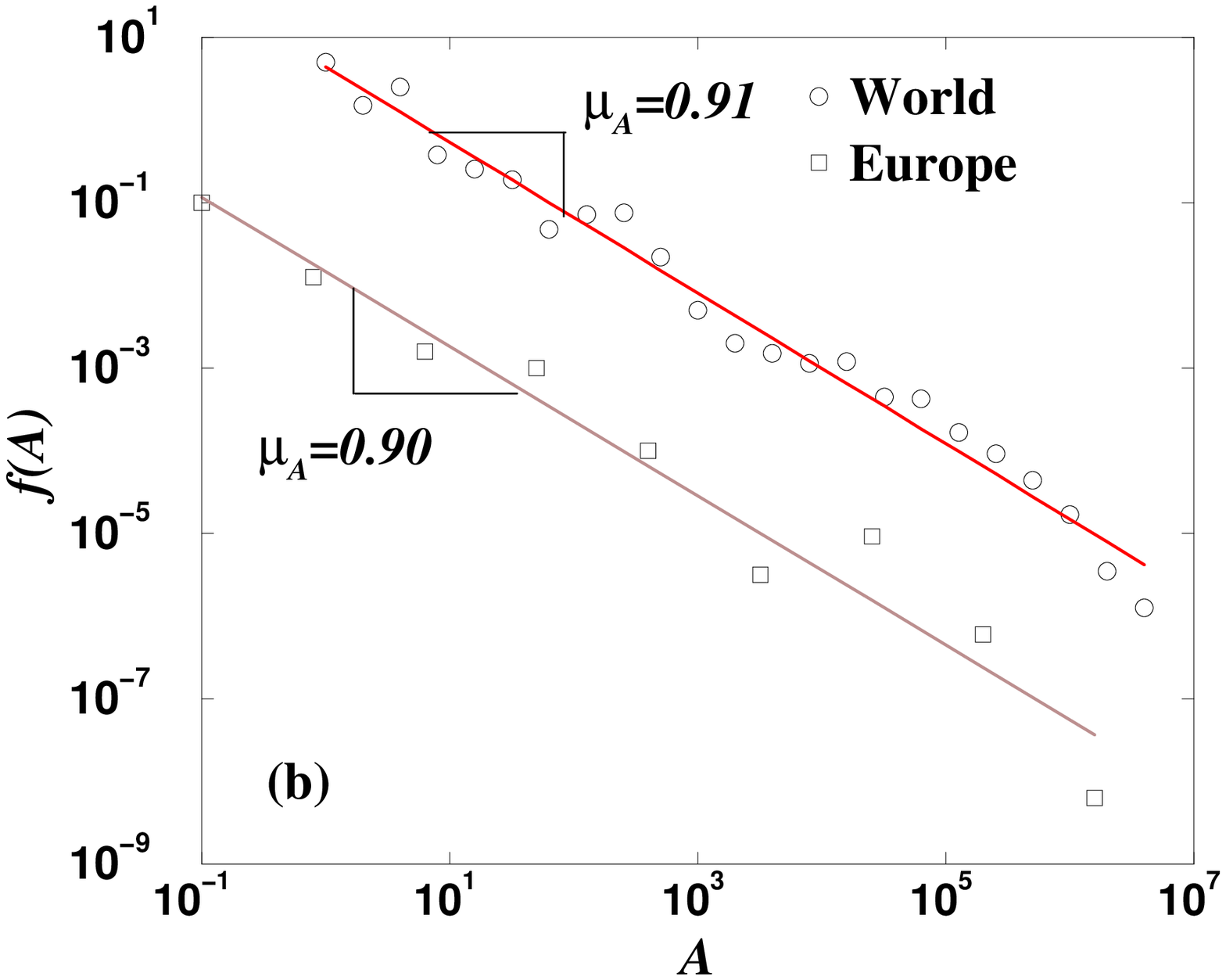}
\end{center}
%\end{minipage}
%\mbox{ \epsfxsize=10cm \epsffile{ fig1a.eps } }
\end{figure}

\begin{figure}
%\begin{minipage}{\dimen0}
\begin{center}
\includegraphics[width=6.0cm]{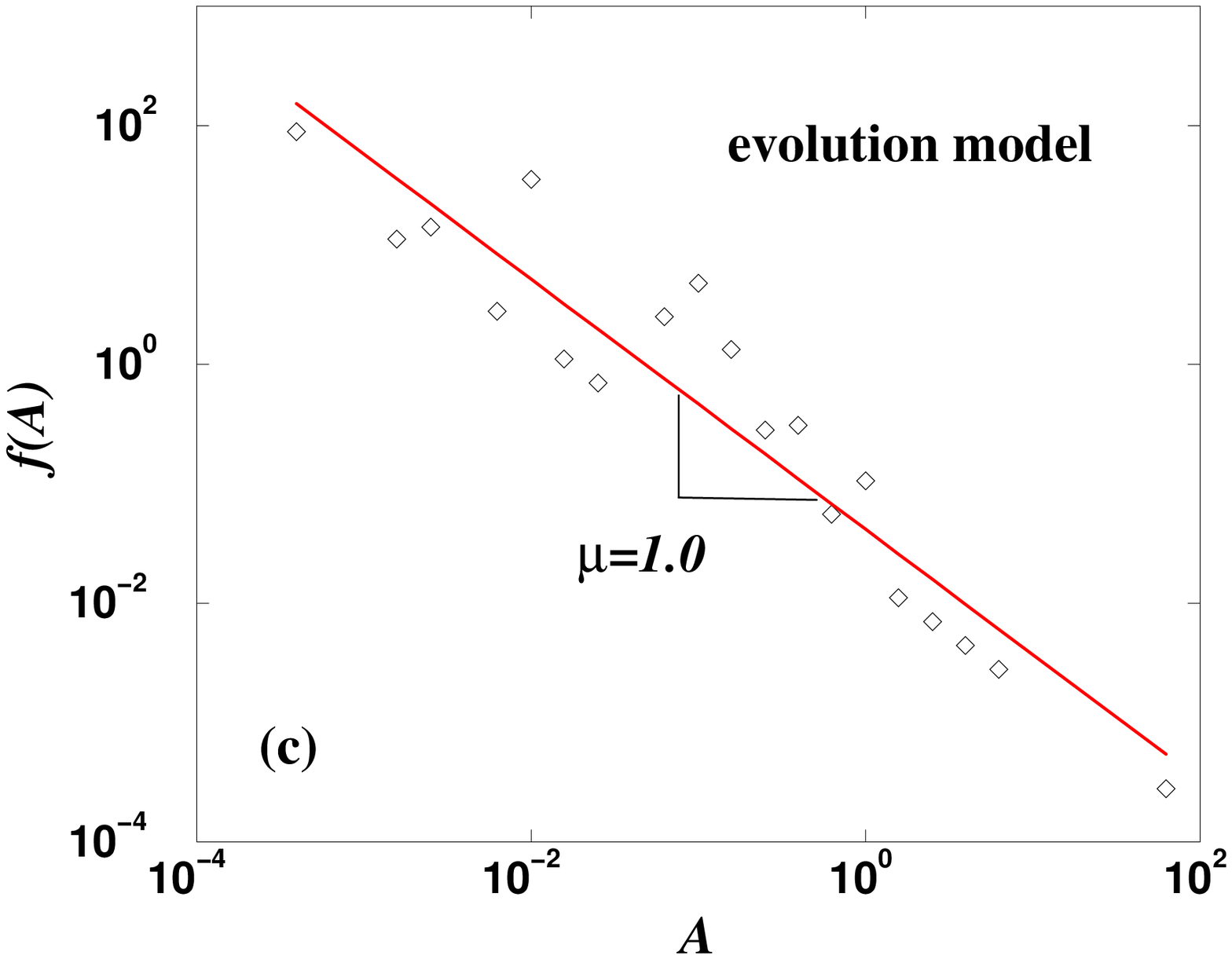}
\end{center}
%\end{minipage}
%\mbox{ \epsfxsize=10cm \epsffile{ fig1a.eps } }
\end{figure}

\begin{figure}
%\begin{minipage}{\dimen0}
\begin{center}
\includegraphics[width=6.0cm]{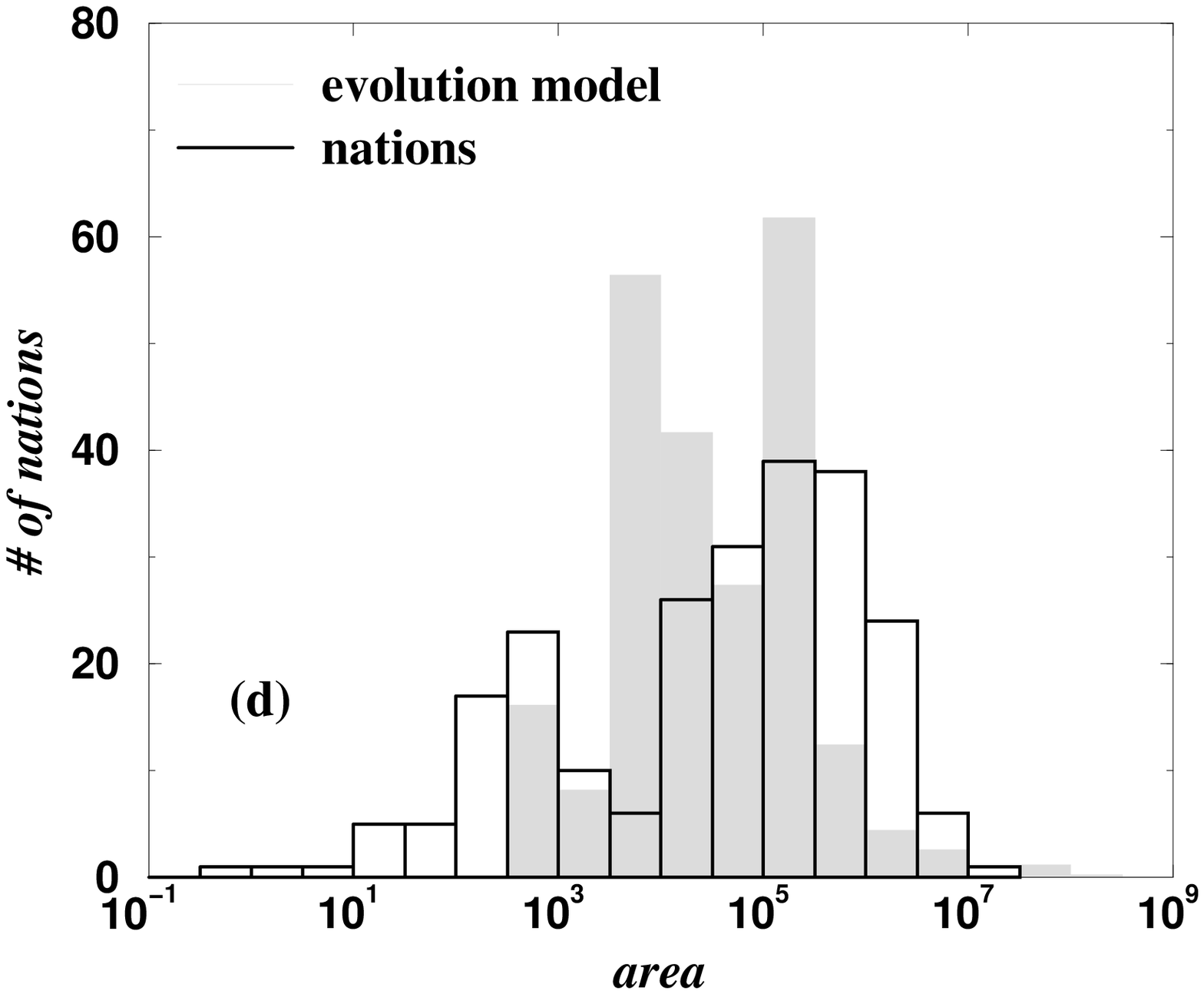}
\end{center}
%\end{minipage}
%\mbox{ \epsfxsize=10cm \epsffile{ fig1a.eps } }
\end{figure}

\begin{figure}
%\begin{minipage}{\dimen0}
\begin{center}
\includegraphics[width=6.0cm]{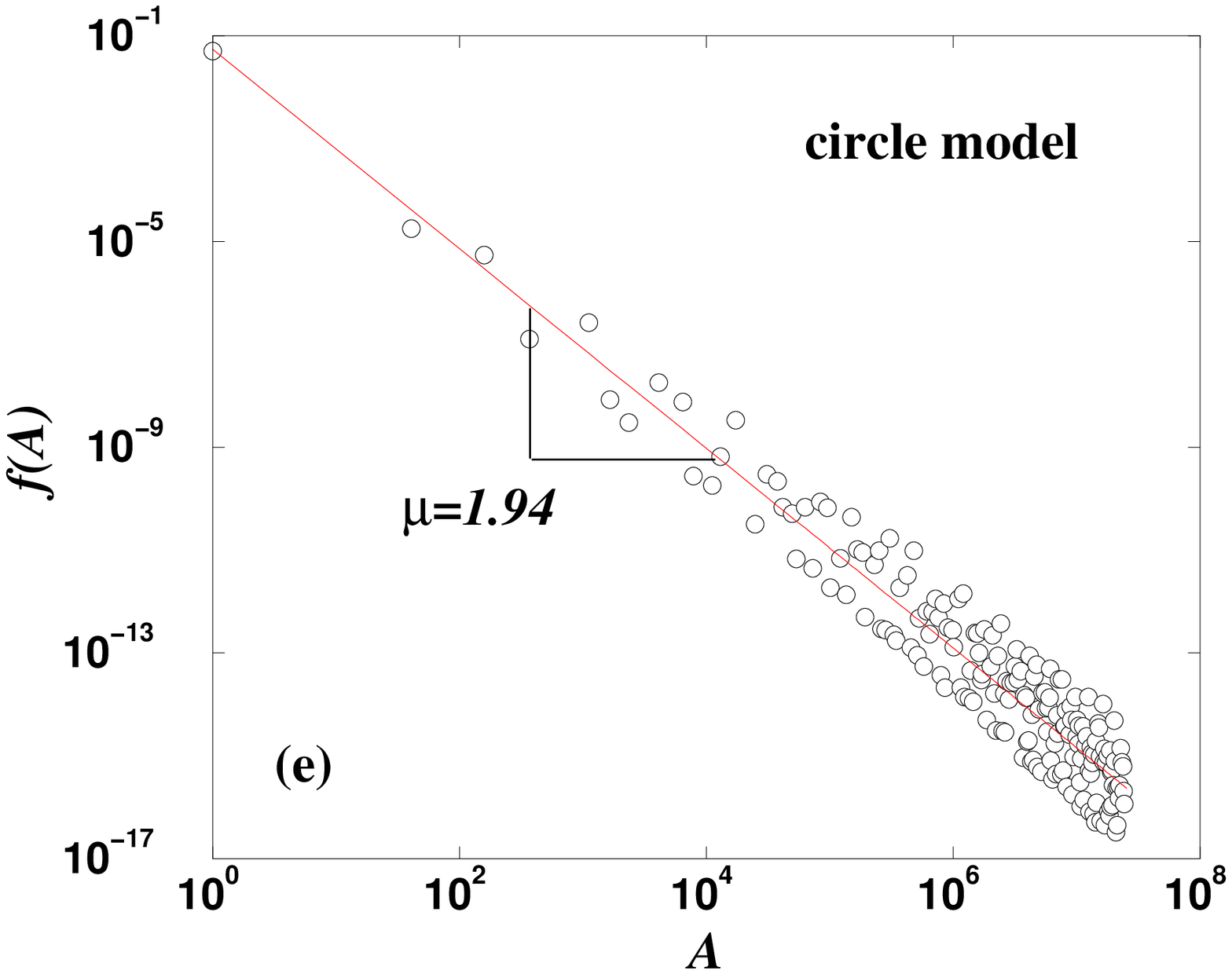}
\end{center}
%\end{minipage}
%\mbox{ \epsfxsize=10cm \epsffile{ fig1a.eps } }
\end{figure}

%\begin{figure}
%\caption{
{\small FIG.1. (a) Double-logarithmic plot of the histogram of the population,
$P$, of the 255 world nations and 140 largest urban agglomerates in the
USA in mid-1997. For nations, the slope gives $\mu_P=0.97$ and the
linear regression coefficient is $R=0.97$; for cities $\mu_P=1.94$ and
$R=0.99$. For visual convenience, city data are multiplied by $10^2$.
(b) Double-logarithmic plot of the histogram of areas, $A$, of the 255
nations of the world and 49 European nations; the corresponding
coefficients are $R=0.99$ and 0.99 respectively.  European data are
divided by $10^2$. The source for both plots is {\tt
http://www.stats.demon.nl}. 
(c) Double-logarithmic plot of the histogram of areas for the 
evolution model.
(d) Distributions of logarithms of nations areas produced by
(i) 255 nations and (ii) by the evolution model.
(e) Double-logarithmic plot of the histogram of areas for the circle 
model; the linear regression coefficient is $R=0.96$.
}
\label{fig:1}
%\end{figure}
%\mbox{ \epsfxsize=10cm \epsffile{ fig1a.eps } }

\end{multicols}
\end{document}